\documentclass[11pt,letterpaper,notoc]{JHEP}
\usepackage{amsmath}
\newcommand{\SM}{ Standard Model}
\DeclareMathOperator{\tr}{tr}
\DeclareMathOperator{\diag}{diag}
\newcommand{\CL}{{\cal L}}
\usepackage{cite}
\title{The Minimal Moose for a Little Higgs}

\author{N. Arkani-Hamed\\ Jefferson Laboratory of Physics, Harvard
  University, Cambridge, MA 02138 \\ email:
  \email{arkani@carnot.harvard.edu}} 
\author{A.G.~Cohen\\ Physics
  Department, Boston University, Boston, MA 02215 \\ email:
  \email{cohen@bu.edu}} 
\author{E. Katz, A.E. Nelson \\ Department of
  Physics, Box 1560, University of Washington, Seattle, WA 98195-1560
  \\ email: \email{amikatz@fermi.phys.washington.edu},
  \email{anelson@phys.washington.edu}}
\author{T. Gregoire, J. Wacker \\ Department of Physics, UC Berkeley,
  Berkeley, CA 94720, USA \\ email:
  \email{gregoirt@socrates.berkeley.edu},
  \email{jgwacker@socrates.berkeley.edu}}

\preprint{BUHEP-02-24\\UW/PT-01/09\\
  HUTP-02/A016} 

\abstract{Recently a new class of theories of electroweak symmetry
  breaking have been constructed. These models, based on
  deconstruction and the physics of theory space, provide the first
  alternative to weak-scale supersymmetry with naturally light Higgs
  fields and perturbative new physics at the TeV scale.  The Higgs is
  light because it is a pseudo-Goldstone boson, and the quadratically
  divergent contributions to the Higgs mass are cancelled by new TeV
  scale ``partners'' of the {\em same} statistics.  In this paper we
  present the minimal theory space model of electroweak symmetry
  breaking, with two sites and four link fields, and the minimal set
  of fermions.  There are very few parameters and degrees of freedom
  beyond the \SM. Below a TeV, we have the \SM\ with two light Higgs
  doublets, and an additional complex scalar weak triplet and singlet.
  At the TeV scale, the new particles that cancel the 1-loop quadratic
  divergences in the Higgs mass are revealed. The entire Higgs
  potential needed for electroweak symmetry breaking---the quartic
  couplings as well as the familiar negative mass squared---can be
  generated by the top Yukawa coupling, providing a novel link between
  the physics of flavor and electroweak symmetry breaking.}

\begin{document}

\section{Introduction}

The \SM\ provides an excellent description of all particle physics
experiments performed to date. The parameterization of electroweak
symmetry breaking in terms of a fundamental scalar field, however, is
almost certainly incomplete.  The quadratically divergent radiative
corrections to the Higgs mass suggest a TeV-scale cutoff. A
more complete, natural theory of electroweak symmetry breaking would
include a stabilization mechanism for the Higgs mass through new
physics at or below the TeV scale.

Precision electroweak measurements are consistent with perturbative
standard model calculations and a light Higgs.  The unreasonable
effectiveness of this fundamental Higgs theory of electroweak symmetry
breaking suggests that any new TeV scale physics that stabilizes the
Higgs mass is also perturbative. A widely held belief is that the only
perturbative candidate for electroweak symmetry breaking that
stabilizes the weak scale is low-energy supersymmetry. In
supersymmetric theories every \SM\ field has a superpartner of
opposite statistics. The quadratic sensitivity of the Higgs mass to
the TeV scale is removed by cancellations of the radiative corrections
of \SM\ fields with those of their superpartners.

Recently, counter to this belief, a different class of models with
perturbative physics stabilizing the electroweak scale has been
introduced \cite{Arkani-Hamed:2001nc,Arkani-Hamed:2002pa}, based on
``deconstruction'' \cite{Arkani-Hamed:2001ca,Hill:2000mu} and the
physics of theory space \cite{Arkani-Hamed:2001ca,
  Arkani-Hamed:2001ed}. In these models the light Higgs field appears
as a pseudo-Goldstone boson at TeV energies and below, with
conventional gauge, Yukawa and self-couplings. The quadratic
sensitivity to the cutoff scale that these couplings normally induce
are cancelled, not by particles of opposite statistics, but instead by
particles of the same statistics. Global symmetries of the theory
ensure these cancellations. The interactions responsible for the
symmetry breaking giving rise to the pseudo-Goldstone bosons are
characterized by a scale $\sim 10$ TeV.

The models described in \cite{Arkani-Hamed:2001nc} are non-linear
sigma models characterized by a toroidal ``theory space''. The small
example described in detail in \cite{Arkani-Hamed:2002pa} has 4 sites
(corresponding to a gauge group $SU(3)^3\times SU(2)\times U(1)$) and
8 links (corresponding to 8 sets of non-linear sigma model fields).
Some of these sigma model fields are eaten, higgsing the gauge group
to the \SM\ group $SU(2)\times U(1)$ and giving TeV scale masses to
the corresponding gauge bosons.  Additional operators break the global
symmetries and give TeV scale masses to most of the scalars, leaving
two pseudo-Goldstone multiplets massless at tree level.  These include
a pair of \SM\ Higgs doublets, interacting through a quartic
potential.  A set of colored, vector-like fermions were introduced to
generate the top quark Yukawa coupling. The additional TeV scale
degrees of freedom cancel the one-loop \SM\ quadratic divergence in
the Higgs mass: the massive gauge bosons cancel the divergence from
the \SM\ gauge loop; the massive scalars cancel the divergence from
the Higgs self-coupling; the massive fermions cancel the divergence
from the top quark loop. Such models are consistent with
precision electroweak constraints (such as the $S$-parameter), and
flavor model building in this context can be explored relatively
unfettered by the bonds of flavor violation.

As emphasized in \cite{Arkani-Hamed:2001nc}, the essential idea of the
Higgs as a pseudo-Goldstone boson and the models constructed are
independent of extra dimensions and their deconstruction.  Models
based on deconstruction have the virtue of a large collection of
approximate symmetries protecting the Higgs mass, and a restricted set
of symmetry breaking effects. The symmetry and field content of these
theories are naturally represented graphically as sites and links, in
a notation sometimes referred to as ``moose'' \cite{Georgi:1986hf} or
``quiver'' \cite{Douglas:1996sw}. This construction makes it evident
that these theories realize a theory space in which symmetry
breaking terms are localized, while the lightest pseudo-Goldstone
bosons are extended objects: non-contractible loops in theory space.
Such extremely light Goldstone bosons, which can only receive mass
from the combined efforts of more than one symmetry breaking term, are
known as ``little Higgses''.  A general analysis \cite{Gregoire:2002a}
reveals that little Higgses are associated with topological properties
of theory spaces: each little Higgs corresponds to an element of the
fundamental group of the theory space; and the little Higgs potential
is obtained from the group relations.

In this paper and its companion\cite{Arkani-Hamed:2002a} we further
abstract the essential features of the little Higgs idea. We construct
models, based on a moose with 2 sites and 4 links, which have the same
low-energy structure as the toroidal theory space models, and are
economical enough to obviate the need for moose notation.  We also
point out a new possibility for the origin of the Higgs potential:
both the Higgs negative mass squared {\em and} the quartic couplings
can have their source in the same operators that generate the top
Yukawa coupling.  This leads to a relation between the Higgs and top
masses $m_H \sim m_t$, and a fascinating link between the physics of
flavor and electroweak symmetry breaking.

The companion paper\cite{Arkani-Hamed:2002a} considers sigma models
based on more general cosets, with no obvious moose description.
Remarkably the Higgs quartic potential in such models can be generated
by the gauge interactions alone. These more general coset
constructions allow the theory below a TeV to contain {\em only} the
\SM\ particles with a single Higgs, and the smallest number of new
states at a TeV.

\section{A Minimal Moose}
\subsection{The Model}
As a warm-up, in this section we present an example of a minimal
moose with only two sites and four links.  It is related to the $2
\times 2$ toroidal construction of \cite{Arkani-Hamed:2001nc}, but is
half as large\footnote{To see the relation, consider a version of the 4
  site toroidal moose with 2 $SU(2)\times U(1)$ sites and 2 $SU(3)$
  sites---the $SU(2)\times U(1)$ sites are at opposite corners. Now
  identify links and sites which are exchanged by a $Z_2$ discrete
  symmetry, or, in other words, orbifold by a translation along the
  diagonal of the moose.}.

The gauge symmetry at one of the sites is $G_1=SU(3)$. (Alternatively
an $SU(2) \times U(1)$ subgroup would suffice.) At the other site the
gauge symmetry is $G_2 = SU(2) \times U(1)$. There are 4 link fields
$X_j$, $j=1,\cdots,4$, which are $3 \times 3$ non-linear sigma model
fields $X_j = \exp(2 i x_j/f)$, each transforming as a
bi-fundamental under $G_1\times G_2$, where a fundamental of $G_2$ is
taken to be 
$\mathbf{2}_{1/6} \oplus \mathbf{1}_{-1/3}$.
All the \SM\ fermions
are charged under the $G_2$ gauge symmetry with their
usual quantum numbers, so the model is anomaly free.  The
theory has a large, approximate $SU(3)^8$ global symmetry
spontaneously broken to $SU(3)^4$, with the non-linear realization
\begin{equation}
  X_j \to L_j X_j R_j^\dagger, \quad j=1,2,3,4 \ .
\end{equation}
The cutoff of this non-linear sigma model is $\Lambda \sim 4 \pi f$,
which we take to be $\sim 10$ TeV. The effective theory beneath this
cutoff is described by the Lagrangian
\begin{equation}
  {\cal L} = {\cal L}_{G} + {\cal L}_{X} + {\cal
    L}_{t} + {\cal L}_{\psi} \ .
\end{equation}
Here ${\cal L}_{G}$ includes the conventional non-linear sigma model
field kinetic terms and gauge interactions, while ${\cal L}_{X}$
contains ``plaquette'' couplings between the $X_j$:
\begin{equation}
  {\cal L}_{X} = f^4 \tr  \left(A X_1 X_2^\dagger 
    X_3 X_4^\dagger \right) +
  f^4 \tr \left(A^\prime X_1
    X_4^\dagger X_3 X_2^\dagger \right) + \text{h.c.}  
\end{equation}
where $A = \kappa \mathbf{1} + \epsilon \mathbf{T}_8$ and $A^\prime
= \kappa^\prime \mathbf{1} + \epsilon^\prime \mathbf{T}_8$.  Each of
these terms breaks the global symmetries, but preserves a sufficiently
large subgroup of the global symmetries to leave some Goldstone bosons
massless. In writing down these couplings, we allow a specific set of
global symmetry breaking operators without addressing their UV origin.
We then include all terms needed to renormalize the theory at 1-loop,
with coefficients which are no smaller than their natural size.
Possible difficulties with a natural origin of the plaquette terms
from QCD-like UV completions of the non-linear sigma model are
discussed in ref. \cite{Lane:2002pe}. In section \ref{top}, we show
that the necessary plaquette terms {\em can} be generated, quite
naturally, from the top sector.

The remaining terms generate the \SM\ Yukawa couplings; to avoid
introducing quadratic divergences, we also introduce a vector-like
pair of colored Weyl fermions $U,U^c$ and couple them to the top
quark with ${\cal L}_t$:
\begin{equation}
  \label{eq:yuk}
  {\cal L}_t = \lambda f 
  \begin{pmatrix} 
    0 & 0 & u_3^{c \prime} 
  \end{pmatrix} 
  X_1 X_2^\dagger 
  \begin{pmatrix} 
    q_3\\ U 
  \end{pmatrix} 
  + \lambda^\prime f U U^c \ .
\end{equation}
Finally, ${\cal L}_\psi$ contains the Yukawa couplings for the light
fermions; since these are very small, the quadratic divergences
associated with them are negligible for our cutoff $\Lambda \sim
10$ TeV. For the light up-type quarks, ${\cal L}_\psi$
has the same form as ${\cal L}_t$ with $U,U^c$ removed, while for
the down and charged lepton sectors ${\cal L}_\psi$ contains
\begin{equation}
  {\cal L}_\psi \supset 
  \begin{pmatrix} 
    q & 0 
  \end{pmatrix} 
  X_1 X_2^\dagger
  \lambda_{\text{D}} f
  \begin{pmatrix} 
    0 \\ 0 \\ d^c 
  \end{pmatrix} +
  \begin{pmatrix} 
    l & 0 
  \end{pmatrix} 
  X_1 X_2^\dagger
  \lambda_{\text{E}} f
  \begin{pmatrix} 
    0 \\ 0 \\ e^c 
  \end{pmatrix} \ .
\end{equation}
This completes the description of the model.

\subsection{Tree-level spectrum and interactions}
Let us examine this theory at tree-level, for the moment putting
$\epsilon = \epsilon^\prime=0$ in ${\cal L}_X$.  The non-linear sigma
model fields Higgs the $G_1 \times G_2$ gauge group down to the $SU(2)
\times U(1)$ subgroup. The massive $SU(3)$ gauge bosons have a mass
$\sim g f$.  The linear combination of Goldstone bosons $(x_1 + \cdots
x_4)$ is eaten, while the quartic terms in ${\cal L}_X$ give mass to
the linear combination $x_1 - x_2 + x_3 - x_4$, of order $\sim \kappa
\ f$. (Here and in what follows we assume $\kappa, \kappa^\prime$ are
real and positive, although it suffices for
$\Re(\kappa,\kappa^\prime)>0$.)  At tree-level two orthogonal
combinations of pseudo-Goldstone boson multiplets, which we can take
to be $x_1 - x_3$ and $x_2 - x_4$, are massless. They decompose under
the unbroken $SU(2) \times U(1)$ gauge symmetry as a pair of Higgs
doublets, as well as a complex weak triplet and singlet.  Furthermore,
these classically massless pseudo-Goldstone bosons, or ``little
Higgses'', receive no 1-loop quadratically cutoff sensitive
corrections to their masses.

At tree-level the little Higgses interact through a quartic
potential, with one linear combination of the doublets coupling to the
top quark. The quartic potential can very quicly be found in the limit
$\kappa = \kappa^\prime$;
with this choice ${\cal L}_X$ has a symmetry under which $X_j \to
X_j^\dagger$, or equivalently $x_j \to -x_j$. This ensures that the
potential is even in the fields, and in deriving the potential for the
light fields, we can simply set the heavy fields to zero. The light fields corresponding to $x_1 - x_3$ and $x_2 -
x_4$ can be parameterized as
\begin{align}
  X_1 = X_3^\dagger = U &\equiv e^{2 i (x+y)/f} \\
  X_4 = X_2^\dagger = V &\equiv e^{2 i (x-y)/f} \ .
\end{align}
The plaquette interactions then give rise to the potential
\begin{equation}
  - \kappa f^4 \tr\left(U V U^{-1} V^{-1}\right) + V \leftrightarrow
  V^{-1} +  \text{h.c.} =
  \kappa \tr \left[x,y\right]^2 + \cdots\ .
\end{equation} 
where we have furhter assumed for simplicity that $\kappa$ is real.

It is extremely interesting that we have found a potential for $x,y$ with
no mass terms but with a quartic potential. Why did this happen? In order
to understand this, first take a limit where one of the
plaquettes is turned off, say by taking $\kappa^{\prime} = 0$. The $\kappa$
plaquette still gives mass to the linear combination $x_1 - x_2 + x_3 -
x_4$, but it is easy to see that no potential can be generated at all for
the other goldstones. This is because the $\kappa$ plaquette still
preserves an $SU(3)^4$ subgroup of the $SU(3)^8$ global symmetry, where 
$R_1=R_2,L_2=L_3,R_3=R_4,L_4=L_1$. This is spontaneously broken to the
diagonal $SU(3)$, leaving three exactly massless goldstone bosons (one of which
is eaten). We can
also see this directly in expanding the plaquette interaction in terms of
the heavy multiplet $z \propto x_1 - x_2 + x_3 - x_4$ as well as the uneaten $x,y$ fields, which yields
schematically 
\begin{equation}
\kappa \tr (f z + i[x,y] + \cdots)^2
\end{equation}
Upon integrating out the heavy $z$ multiplet there is no potential for
$x,y$. Diagramatically, there is a quartic coupling $\tr [x,y]^2$, as well as a
cubic coupling $\tr z[x,y]$. Integrating out $z$ exactly cancels the
quartic coupling, as it must since $x,y$ are exact Goldstone bosons. An
exactly analagous argument holds for the $\kappa^{\prime}$ coupling. However,
$\kappa^\prime$ preserves a {\it different} $SU(3)^4$ global symmetry, and the
potential from the $\kappa^\prime$ plaquette is of the form
\begin{equation}
\kappa^\prime \tr (fz - i[x,y] + \cdots)^2
\end{equation}
In the presence of both $\kappa$ and $\kappa^\prime$, there is only an $SU(3)
\times SU(3)$ global symmetry broken to $SU(3)$, and only one exactly
massless goldstone boson, which is the one that is eaten. Therefore, $x,y$ are not
exact Goldstone bosons and can acquire a potential. However, since their
potential is the sum of two pieces, one proportional to $\kappa$ and the
other proprtional to $\kappa^\prime$, it is impossible for $x,y$ to pick up
a mass at tree level. They {\it can} acquire a quartic potential, however: 
upon integrating out $z$ we have 
\begin{equation}
\frac{\kappa \kappa^\prime}{\kappa + \kappa^\prime} \tr [x,y]^2
\end{equation}
Note the non-analytic dependence on $\kappa,\kappa^\prime$ in the
denominator, arising from integrating out $z$ which has a mass squared proprtional
to $f^2 (\kappa + \kappa^\prime)$. This quartic coupling vanishes as it
must in the limit where either $\kappa$ or $\kappa^\prime$ vanishes.  
Thus we have generated a quartic potential for the little Higgses, 
without a mass term, by breaking the global symmetry with two different
couplings. Any one of these couplings preserves enough global symmetry to
ensure that the little Higgses are
exact Goldstone bosons. But together the couplings break all these
symmetries and the little Higgses can acquire a quartic potential. 
In the next section, we see that this same
mechanism ensures the absence of 1-loop quadratic divergences for the mass
of the little Higgses. 

We can exhibit the components of $x$ as a $3 \times 3$ hermitian
matrix
\begin{equation}
  x = 
  \begin{pmatrix} 
    \varphi_x + \eta_x &
    h_x \\ h_x^{\dagger} & -2 \eta_x
  \end{pmatrix}
\end{equation}
and similarly for $y$. Here $\varphi,\eta$ are fields in the
$\mathbf{3}_0, \mathbf{1}_0$ representation of $SU(2) \times U(1)$
respectively, while the $h$ have the quantum numbers
$\mathbf{2}_{1/2}$ of the standard model Higgs. The quartic potential
is then 
\begin{equation}
  \kappa \tr \left[x,y\right]^2 = \kappa 
  \tr (h_x h_y^{\dagger} - h_y h_x^{\dagger})^2
  + \kappa (h_x^{\dagger} h_y - h_y^{\dagger} h_x)^2 + \text{ terms
    involving }   \varphi, \eta\ .
\end{equation}
This can be recast in a more familiar form by defining $h_1 = h_x
+ i h_y, h_2 = h_x - i h_y$. The quartic potential is then
\begin{equation}
  \kappa \tr (h_1 h_1^{\dagger} - h_2 h_2^{\dagger})^2
  + \kappa (h_1^{\dagger} h_1 - h_2^{\dagger} h_2)^2 + \text{ terms
    involving }   \varphi, \eta\ .
\end{equation}
This two-Higgs doublet quartic potential is similar to that of the
supersymmetric \SM.

To exhibit the top Yukawa coupling, we expand ${\cal L}_t$ to
first order in the Higgs doublet fields
\begin{equation}
  \lambda  u_3^{c \prime} \left(f U + h_x q_3 \right) + \cdots +
  \lambda^\prime f U U^c 
\end{equation}
One linear combination of $U^c$ and $u_3^{c \prime}$ marries $U$ to
become a massive fermion with mass $\sim \lambda f$; the orthogonal
combination $u_3^c$ remains massless with a Yukawa coupling to $q_3$
\begin{equation}
  \lambda_t u_3^c h_x q_3, \quad \text{where} \quad \lambda_t =
  \frac{\lambda \lambda^\prime}{\sqrt{\lambda^2 + \lambda^{\prime 2}}}
\end{equation}
The mixing of the top quark with vector-like Fermions at the TeV
scale is similar to Frogatt-Nielsen models of flavor
\cite{Froggatt:1979nt} and the top see-saw \cite{Dobrescu:1998nm,
  Chivukula:1998wd}.

In summary, at the classical level there are two massless Higgs
doublets, together with a complex triplet and singlet. These scalars
have a tree-level quartic potential, and one linear combination of the
Higgs doublets has a Yukawa coupling to the top quark fields $q_3,
u_3^c$. We also have a set of massive vectors, scalars and fermions
with masses $\sim gf, \kappa f, \lambda f $ respectively. All these
scales are of order a TeV.

\subsection{Power-counting and absence of 1-loop quadratic divergences}
\label{absence}
Radiative corrections generate masses for the little Higgses that are
only logarithmically sensitive to the cutoff.  We establish this
through two different routes. First we examine how the non-linearly
realized symmetries which protect the little Higgs masses are explicitly
broken: we show that any one of the gauge, plaquette or Yukawa
interactions alone preserve enough of these symmetries to forbid
masses for the Goldstone multiplets.  Any quadratically divergent
correction to the masses must then arise from a combination of more
than one of these couplings, and is absent at 1-loop. Secondly we
give a simple, general set of rules which are sufficient (though not
necessary) to ensure an arbitrary theory space to be free of 1-loop
quadratic divergences, and verify the validity of these rules by
directly computing the 1-loop Coleman-Weinberg potential.  Our model
can trivially be seen to satisfy these rules.

The nonlinearly realized $SU(3)^8$ symmetry is explicitly broken by
the gauge, plaquette and Yukawa interactions, and these will in turn induce
other operators. We do a standard power-counting analysis
\cite{Manohar:1984md, Cohen:1997rt, Luty:1998fk}
in order to determine the natural size of these
interactions. This is most straightforwardly done
following \cite{Cohen:1997rt, Luty:1998fk}. The Lagrangian is written as ${\cal L} = \Lambda^4/16
\pi^2 \hat{{\cal L}}$, where $\Lambda \sim 4 \pi f$ is the UV cutoff and
all the mass scales in $\hat{{\cal L}}$ are scaled by powers of
$\Lambda$. This rule leads to the familar $f^2$ coefficient for the
goldstone kinetic terms. Also all non-derivative terms involving the $X$'s 
naturally have a
coefficient $\Lambda^2 f^2$, while the fermion mass terms and coupling to
$X's$ are scaled by
$\Lambda$. The small symmetry breaking effects of the spurions in ${\cal
L}_X$ then have small dimensionless size $\sim
\kappa^{(\prime)}/(16 \pi^2), 
\epsilon^{(\prime)}/(16
\pi)^2$, while those in ${\cal L}_t$ are $\sim
\lambda^{(\prime)}/(4 \pi)$. Note that we can independently rephase
$\lambda$ and $\lambda^\prime$ by rephasing the fermions $u^c$ and
$U^c$, therefore in any induced operator only involving the $X$'s, 
these spurions can only enter as $|\lambda^{(\prime)}/4 \pi|^2$. 
The gauge interaction spurion is $g^2/(16 \pi^2)$. Therefore each one of
our spurions counts as a loop factor. 

The power-counting is
now staightforward. Every operator is proportional $\Lambda^2 f^2$ times
the appropriate product of spurions needed to generate it. In particular,
any induced little Higgs masses are proprtional to $\Lambda^2$ times the product of
spurions. 
The generation of little Higgs masses will require at least
2 spurions, and therefore will have quadratic sensitivity to
the cutoff only at 2-loop level. This happens because each one of our interaction
terms preserves a large subset of the global symmetry. Consider
the limit where only one of the gauge couplings, say the $G_1$ gauge
coupling, is non-zero. This coupling preserves a symmetry under which
$L_1 = L_2 = \cdots =l$ but all four of the $R_j$ arbitrary. One
combination of the Goldstone multiplets is eaten, but  {\em
three} sets of  Goldstone modes remain massless. Exactly the same happens
when only the couplings of $G_2$ are non-zero. The presence of both
couplings breaks all the chiral symmetries and there are no exact
Goldstone modes. However {\em two} sets are left classically massless.
Any quadratically divergent mass must involve both couplings, arising
at 2-loop order. A similar analysis applies to the plaquette
interactions. As we discussed in the last section, in the limit where only $\kappa$ is non-zero, the global
symmetry is the $SU(3)^4$ subgroup of $SU(3)^8$ with
$R_1=R_2,L_2=L_3,R_3=R_4,L_4=L_1$. This is spontaneously broken to the
diagonal $SU(4)$, leaving three massless Goldstone multiplets.
With only $\kappa'$ non-zero a different $SU(3)^4$ is broken, while in
the presence of both $\kappa$ and $\kappa'$ only an $SU(3) \times
SU(3)$ symmetry is left, with the one exact Goldstone boson eaten via
the Higgs mechanism.  The remaining two approximate Goldstone
multiplets only acquire quadratically divergent masses if both
$\kappa$ and $\kappa'$ are present, again requiring at least 2 loops.
Finally a similar analysis applies to the fermions. The
addition of $U$ to the $q_3$ gives the $\lambda$ piece of
${\cal L}_t$ an $SU(3)$ global symmetry, which ensures that $h_x$ is a
Goldstone mode. This symmetry is broken by the mass term
$\lambda^\prime f U U^c$. But again this requires both $\lambda$ and
$\lambda^\prime$, and arises only at 2-loop level. Note that it is
important here that there are rephasing symmetries that force any one of
our spurions to appear quadratically as $|\lambda^{(\prime)}/(4 \pi)|^2$. This
would not be the case if we add a mass term $\lambda^{''} f u_3^{c \prime} U$. 
Then, the combination $(\lambda^*/4 \pi)(\lambda^{''}/4 \pi)$ is invariant,
and would lead to a 1-loop quadratically divergent Higgs mass. However this
$u_3^{c \prime} U$ mass term can be prohibited by chiral symmetries. 

There are simple rules which are sufficient (although not necessary)
to ensure the absence of 1-loop quadratic divergences for the little
Higgs mass from the gauge, quartic and top Yukawa sectors in a general
moose.  For the gauge and quartic couplings these rules are simply
phrased as properties of the theory space:
\begin{description}
\item[Gauge couplings:] Every link connects two different sites.
\item[Quartic couplings:] No plaquette contains the same link twice.
\end{description}
There are a variety of ways to ensure the absence of 1-loop quadratic
divergences from the top sector. The simplest possibility is just what
we have done for ${\cal L}_t$, but which we phrase here slightly  more
generally:
\begin{description}
\item[Top Yukawa couplings:] The top Yukawa comes from interactions of
  the form
\begin{equation}
  \lambda f
  \begin{pmatrix} 
    0 & 0 & u_3^{c \prime} 
  \end{pmatrix} 
  W 
  \begin{pmatrix} 
    q_3\\ U
  \end{pmatrix} 
  + \lambda^\prime f U U^c
\end{equation}
\end{description}
where $W$ is a product of link fields.

Note that our specific 2-site, 4-link theory space satisfies all of
these properties: each link connects the two sites, the plaquette
interactions in ${\cal L}_X$ contain each link only once, and ${\cal
  L}_t$ has precisely the form of our rule.  This also makes it clear
why the minimal model has 2 sites and 4 links.  These rules are a
manifestation of the general requirement that all order one symmetry
breaking terms must preserve at least one global symmetry under which
the little Higgses transform nonlinearly. Any order one couplings for
the little Higgses must arise from a collaboration between at least
two such symmetry breaking terms.

To show that these rules suffice to eliminate 1-loop quadratic
divergences to the little Higgs masses, we compute the
quadratically divergent part of the 1-loop Coleman-Weinberg potential.
We turn on a background $X_i = \tilde{X}_i$ for the link fields. The
1-loop quadratic divergences in the Coleman-Weinberg potential are
proportional to
\begin{equation}
  \frac{\Lambda^2}{16 \pi^2} \text{Str} M^\dagger M[\tilde{X}]\ .
\end{equation}
We must therefore calculate $\text{Str} M^{\dagger}M[\tilde{X}]$ where
$M[\tilde{X}]$ is the mass matrix of the theory in the presence of the
background.

First consider the gauge sector.  Consider a link field between two
different sites $i$ and $j$.  The gauge boson mass matrix comes from
expanding the covariant kinetic term for the link fields to quadratic
order in the gauge fields, yielding $A_i^a (M^2[\tilde{X}])^{ij}_{ab}
A_j^b$, where $a,b$ are gauge indices and
\begin{eqnarray*}
  M^2_{ab}[\tilde{X}] = \frac{f^2}{4}
  \left( \begin{array}{cc}
      \frac{1}{2} g_i^2 \delta_{ab}& g_i g_j m^2_{ab} \\
      g_i g_j  {m^2}_{ab}^\dagger
      & \frac{1}{2} g_j^2 \delta_{ab}
    \end{array}\right) \hspace{0.6in}
  {m^2}_{ab}=\tr \mathbf{T}_a \tilde{X} \mathbf{T}_b
  \tilde{X}^\dagger 
\end{eqnarray*}
The important point is that $M^2_{aa}$ is always independent of the
background field, $\tilde{X}$, and therefore so is the trace.  Hence
there are no 1-loop gauge quadratic divergences for any link field
mass.  This argument breaks down if a link field connects a site to
itself.

Now consider the 1-loop corrections involving the quartic couplings,
which arise from the plaquette interactions. Consider a general 
plaquette:
\begin{eqnarray}
  V(X_i) = - \kappa f^4 \tr M_1 X_1 \cdots M_N X_N + \text{h.c.}
\end{eqnarray}
where the $M_i$ are arbitrary matrices.  Write the link fields as a
linearized fluctuation, $x_i \equiv x_i^a \mathbf{T}_a$, about a
background field, $\tilde{X}_i$: \text{$X_i = \exp(i x_i)
  \tilde{X}_i$.} In this decomposition, the kinetic terms for $x^a_i$
are independent of the background field, which greatly simplifies the
analysis.  We expand the plaquette to quadratic order in the
fluctuations and find the mass matrix, $x_i^a (M^2)_{ab}^{ij} x_j^b$.
The diagonal component of this mass matrix is
\begin{eqnarray}
  (M^2)_{ab}^{ii}
  \sim \kappa f^2 \tr M_1 \tilde{X}_1 \cdots M_i \mathbf{T}_a \mathbf{T}_a
  \tilde{X}_i
  M_{i+1} \tilde{X}_{i+1} \cdots M_{N} \tilde{X}_{N} + \text{h.c.}
\end{eqnarray}
Summing over the diagonal entries of the mass matrix and using $\sum_a
\mathbf{T}_a ^2 \propto \mathbf{1}$, we find $ \tr M^2 \propto \tr M_1
\tilde{X}_1 \cdots M_i \tilde{X}_i M_{i+1} \tilde{X}_{i+1} \cdots
M_{N}\tilde{X}_{N}$ + h.c., which is just the plaquette operator
itself! Therefore, the 1-loop quadratic divergences only renormalize
the plaquette interactions, and do not generate any new operators in
the theory. If a field appears in a plaquette more than once, then
this argument breaks down: the mass matrix will have a more
complicated form with $X$-dependent diagonal entries.

Let us finally check that the absence of 1-loop quadratic divergences from
the Yukawa couplings. For this purpose, it is enough to consider only
the interaction proportional to $\lambda$. The mass matrix for the
relevant fermions in a general background $\tilde{W}$ is then
\begin{equation}
  M = \lambda f P \tilde{W}, \, \, \text{ where }
  P = \diag(0,0,1) \ .
\end{equation}
The quadratic divergence is then proportional to 
\begin{equation}
  \tr M^\dagger M = \lambda^2 f^2 \tr \tilde{W}^\dagger P P \tilde{W}
  = \lambda ^2 f^2 \tr P 
\end{equation}
which is independent of the background $\tilde{W}$. Once again, there
are no 1-loop quadratic divergences. Note that the presence of the
$U$ fields is crucial here. In its absence, we would have instead
$M = \lambda f P \tilde{W}(1 - P)$, and $\tr M^\dagger M$ would indeed
depend on the background field $\tilde{W}$. This is of course a direct
consequence of our spurion analysis. In the absence of
$\lambda^\prime$, the Yukawa sector has an enhanced $SU(3)$ global
symmetry acting on $W$ and the $(q_3,U)$ triplet, and no potential
for the components of $W$ can be generated.

We have verified the absence of quadratic divergences explicitly by
seeing that the trace of the mass squared matrix for the fields in the
theory is independent of the background little Higgs fields. In terms
of mass eigenstates, this means that as the little Higgs background is
turned on, the classically massless modes become heavier, but the
heavy modes become lighter in just such a way that the sum of the
mass squareds is independent of the background. Thus the cancellation
of quadratic divergences is between particles of the {\em same}
statistics: the massive gauge bosons cancel the quadratic divergences
associated with the massless \SM\ gauge fields, the massive scalars
cancel the quadratic divergence associated with the little Higgs
quartic coupling, and the massive fermion cancels the quadratic
divergence from the top-Yukawa coupling.

\subsection{Electroweak symmetry breaking}
While there are no quadratically divergent corrections to the masses
of the little Higgses, in this model there are logarithmically
divergent corrections at 1-loop, and quadratically divergent 2-loop
corrections.  In the general model of \cite{Arkani-Hamed:2001nc}, the
quadratic divergences can be pushed to $N+1$ loops. However, there are
always {\em finite} 1-loop corrections to the Higgs mass squared,
which are at least of order $3 \lambda_t^4 f^2/(16\pi^2)$ and so we
expect $f \sim$ TeV. Since the cutoff can not go far above $4 \pi f
\sim 10$ TeV, removing quadratic divergences beyond 1-loop is an
unnecessary extravagance. In order to compute the magnitude and sign
of the radiative corrections to the little Higgs masses in our model,
we look at the 1-loop contribution, which is slightly log enhanced.

The gauge and quartic couplings give a positive mass squared to all of
$\phi_{x,y},\eta_{x,y},h_{x,y}$. It is easy to understand this
qualitatively: the usual quadratic divergences of the low energy
theory are cut off at the mass of the heavy field which cancels the
divergence. For instance, the gauge loops generate a mass squared $\sim
[g^2/(16 \pi^2)](gf)^2$. The fermion loop does not generate any
potential for $\phi_{x,y},\eta_{x,y}$, since the interactions in
$\CL_t$ do not break the $SU(2)\times U(1)$ chiral symmetries under
which the $\phi$ and $\eta$ fields transform nonlinearly. However, the
fermion loop does produce a negative mass squared $\sim - [3
\lambda^2/(16 \pi^2)] (\lambda f)^2$ for $h_x$, which can dominate over
the positive gauge and plaquette contributions, and so we can have
$m_{h_y}^2 > 0$ while $m_{h_x}^2 < 0$. This forces $h_x$ to acquire a
vev. However the quartic potential has flat directions where $h_x$
has any value while $h_y=0$, and therefore $h_x$ runs away along this
flat direction. The flat direction analysis is perhaps more familiar
in the $h_1,h_2$ basis, where our quartic potential has the same
form as in the supersymmetric standard model. In this basis, the Higgs
mass terms are of the form $(m_x^2 + m_y^2) (|h_1|^2 + |h_2|^2) +
(m_x^2 - m_y^2)h_1^\dagger h_2 + \text{h.c.}$ The familiar flat direction
where $h_1 = h_2$ is not stabilized since $m_{h_1}^2 = m_{h_2}^2$.
This is why we have added the $\epsilon,\epsilon^\prime$ terms to ${\cal
  L}_X$, where $\epsilon$ can be naturally small. Expanding these
pieces to quadratic order generates the mass term $\Im(\epsilon^\prime -
\epsilon) f^2(|h_1|^2 - |h_2|^2)$, which splits $h_1,h_2$ and lifts
the flat direction, stabilizing electroweak symmetry breaking. 

We have seen that EWSB can arise naturally in this very simple model
with a light Higgs, avoiding 1-loop quadratic divergences, and
allowing for a cutoff $\Lambda \sim 10$ TeV.

\subsection{Precision Electroweak and FCNC constraints}

The bounds from precision electroweak data can be satisfied, since all
the new physics in these models decouples in the large mass limit. There
are decoupling effects that can nevertheless be close to experimental
bounds. For instance, the expansion of the non-linear sigma model kinetic
terms can give rise to operators that violate custodial $SU(2)$ and
generate a contribution to the $\rho$ (or $T$) parameter. 
Also, the triplet fields $\phi$ can acquire vacuum
expectation values, after electroweak symmetry breaking, from cubic
terms of the form $h \phi h$, and contribute to the $T$ parameter. Such couplings may arise at
tree-level for the heavy triplets, or at 1-loop for the light
triplets. However, all of these correction to $m_W/m_Z$ are of order 
$\sim (v/f)^2$,
parametrically the same size as 1-loop \SM\ corrections, and give a
correction to $T$ close to the bound.  Another potential problem is
that a cutoff-sensitive tadpole might be generated for the light
singlets, which would lead to a singlet vev of order $f$ and
destabilize the desired ground state. In \cite{Arkani-Hamed:2002pa} a
$Z_4$ global symmetry was imposed which forbids such a tadpole and
also makes this singlet a stable dark matter candidate. However the
$Z_4$ symmetry is not necessary simply to eliminate the tadpole. We
have seen directly that no tadpole is generated at 1-loop: the
plaquette interactions do not contain linear terms, and the fermion
loop does not generate any potential for the triplets or singlets.
More generally, the generation of any such tadpole requires both
interactions which break the global diagonal $SU(3)$ and interactions
which break the nonlinearly realized $ U(1)$ subgroups of the
$SU(3)^8$ under which the $\eta'$s transform nonlinearly.  The
$\kappa,\kappa^\prime$ terms in $ \CL_X$ preserve the $SU(3)$, while
$\CL_G$ or $\CL_t$ preserve all the $U(1)'$s. Any tadpole must involve
at least $\kappa,\kappa^\prime$ and one of the other couplings, and is
small enough that any resulting singlet vev is of order $M_W$ or less.
An $\eta$ tadpole is also forbidden by $CP$, since the $\eta$'s are
$CP$-odd. Therefore, if $\kappa,\kappa^\prime$ are real, no $\eta$ tadpole
is ever generated from this sector of the theory. 
The $\epsilon,\epsilon^\prime$ terms break both the $SU(3)$ and some of
$U(1)$'s, as well as $CP$. Any tadpole involving one of these couplings must further
involve at least a $\kappa$ or $\kappa^\prime$, again to leading to an
acceptably small tadpole. 
 
Another challenge for TeV extensions of the \SM\ is ensuring
sufficient suppression of flavor changing neutral currents (FCNC). Any
flavor changing neutral currents mediated by particles in this
effective theory are smaller than those of the \SM. Furthermore, our
cutoff is high enough to make most FCNC from physics above the cutoff
sufficiently small.  There are a few significant constraints on flavor
physics and the UV origin of $\CL_\psi$, particularly from kaon $CP$
violation and $D-\bar D$ mixing\cite{Chivukula:2002ww}.  As an
existence proof that FCNC are not necessarily a problem, we note that
this moose is easily UV completed into a renormalizable supersymmetric
theory with supersymmetry breaking scale of order the cutoff
\cite{Katz:2002a} which allows the couplings needed for quark and
lepton masses while satisfying FCNC constraints. An alternative is a
``cascade'' theory, in which the nonlinear sigma model is UV completed
into a linear sigma model whose sigma field is itself a little Higgs
of a nonlinear sigma model with a cutoff on the order of hundreds of
TeV---high enough that FCNC from beyond are not a problem.

\section{Higgs potential from Top Yukawa}
\label{top}
In this section, we construct a model where {\em all} of the
interactions needed to generate the Higgs potential---both the negative
mass squared and the plaquette interactions producing the quartic
interactions---arise from the same couplings that generate the top
Yukawa coupling. The site and link structure is exactly the same as
before. We add two sets of vector-like triplet Fermions to the theory,
$T_{1,2},T^c_{1,2}$, where $T_i=(Q_i,U_i), T^c_i = (Q^c_i,U^c_i)$. 
The Lagrangian is of the form 
\begin{equation}
  {\cal L} = {\cal L}_G + {\cal L}_{\psi} + {\cal L}^\prime_t + \cdots
\end{equation}
Note that we don't have any ``plaquette'' interactions analagous to ${\cal L}_X$;
neverthelss, we will see that these interactions must be included with the
naturally correct size, induced from those in the Yukawa interactions of ${\cal
L}^\prime_t$. ${\cal L}^\prime_t$ is of the form 
\begin{equation}
{\cal L}_t^\prime = {\cal L}_{T} + {\cal L}_{mix}
\end{equation}
Here ${\cal L}_T$ represents the interaction of the vector-like $T,T^c$
with the $X's$ as 
\begin{equation}
{\cal L}_t^\prime = f T_1^c W_1 T_1 + f T_2^c W_2 T_2 
\end{equation}
where
\begin{equation}
W_1 = \lambda_1 X_2 X_1^\dagger + \tilde{\lambda}_1 X_3 X_4^\dagger, \quad 
W_2 = \lambda_2 X_4 X_1^\dagger + \tilde{\lambda}_2 X_3 X_2^\dagger
\end{equation}
while ${\cal L}_{mix}$ are mass mixing terms between  $T,T^c$ and the
Standard Model top sector $q_3^\prime,u_3^{c \prime}$
\begin{equation}
{\cal L}_{mix} = 
f q_3^\prime (\zeta_1 Q^c_1 +
\zeta_2 Q^c_2) + f u_3^{c \prime} (\tilde{\zeta}_1 U_1 + \tilde{\zeta}_2 U_2)
\end{equation}
Note that the form of ${\cal L}_T$ violates the ``Yukawa coupling'' rule of
the previous section, and there will indeed be 1-loop quadratic divergences
in this theory. However, it is easy to see that these quadratic divergences
do not generate masses for the little Higgses; instead they generate the
plaquette interaction of the previous section! This is easy to see both
from power-counting, and also directly from the quadratically divergent part
of the Coleman-Weinberg potential, which is proportional to 
\begin{equation}
  \frac{1}{16 \pi^2} \Lambda^2 f^2 \tr \left(|W_1|^2 + |W_2|^2\right) 
  \sim \mbox{const} + f^4 \lambda_1^* \tilde{\lambda}_1 \tr X_1
  X_2^\dagger X_3 X_4^\dagger 
  + f^4 \lambda_2^* \tilde{\lambda}_2 \tr X_1 X_4^\dagger X_3 X_2^\dagger
\end{equation}
Therefore the presence of the Yukawa interactions {\it requires}
plaquette interactions with a natural size $\sim \lambda^2$. It is
easy to extend our power-counting analysis to conclude that
quadratically divergence little Higgs masses are only generated at
2-loop order as before. 
In particular, the operators
\begin{equation}
 \epsilon f^4 \tr {\mathbf T}_8 X_1 X_2^\dagger X_3 X_4^\dagger +
\epsilon^\prime f^4 \tr {\mathbf T}_8 X_1 X_4^\dagger X_3 X_2^\dagger
\end{equation}
are generated; since they require $SU(3)$ breaking they have a natural size 
$\sim \lambda^2 g^2/(16 \pi^2)$ or $\sim  \lambda^2 \zeta^2/(16 \pi^2)$,
and will in general have non-zero imaginary parts. 
Therefore we 
also generate the $\epsilon$ plaquettes needed in the previous section,
with the correct natural size! 
Due to the presence of these terms, all flat directions can be lifted an 
electroweak symmetry breaking can be triggered and stabilized.
It is also straightforward to check that 
that any tadpole for the $\eta$'s can only
arise at 3-loop order or higher, and is sufficiently small.

After mixing with the Standard Model fermions, at low energies we have two
massless Weyl fermions $q_3,u_3^c$ with a Yukawa couplings to a linear
combination of little Higgses 
\begin{equation}
q_3 (\alpha h_x + \beta h_y) u_3^c
\end{equation}
which give rise to top quark mass after electroweak symmetry breaking. 

Note that the plaquette interactions have a natural size $\sim \lambda^2$ which is
parametrically $\sim \lambda_t^2$. These give rise to a Higgs quartic
potential which is $\sim \lambda_t^2$, and therefore in this model the physical
Higgs mass is parametrically $m_H \sim m_t$.

\section{Conclusions}

In this paper we have presented a model of electroweak symmetry
breaking accomplished by a naturally light Higgs scalar. The Higgs
particle is a pseudo-Goldstone boson, and its mass is therefore
protected against large radiative corrections.  The technology of
theory space is useful in constructing general models of this kind,
eliminating sensitive dependence on short distance physics to
arbitrary loop order. Since in constructing models of electroweak
symmetry breaking this dependence need only be postponed to a scale of
tens of TeV, the extremely simple model presented here is entirely
suitable as a realistic theory of electroweak symmetry breaking.

At the TeV scale the physics is perturbative and well described by an
effective Lagrangian with a small number of parameters.  There are
only a small number of new states beyond the \SM. Counting {\em all}
the helicity states, we introduce 56 new states beyond those of the
\SM. As a point of comparison, the supersymmetric standard model
introduces 126 new states.  More importantly, the essential features
of our model are characterized by very few parameters beyond the \SM.
Electroweak symmetry breaking can be triggered through the top quark
couplings, much as in the MSSM. There is some freedom in how the top
quark couplings are incorporated, leading to the possibility that the
entire Higgs potential can arise as a radiative effect at low
energies. In this case the Higgs mass is naturally of the same size as
the top quark mass.

In this effective theory, flavor changing interactions are only
induced through the operators which give rise to the fermion Yukawa
couplings, and dangerous flavor changing effects (often associated
with physics beyond the standard model) do not arise. As always in an
effective theory, there is the possibility of flavor changing neutral
currents from physics above the cutoff, but it is straightforward to
conceive scenarios where this also is not a problem.

For twenty years the domain of perturbative electroweak symmetry
breaking models with a naturally light Higgs has been ruled by
supersymmetric theories.  In the last year a viable challenger has
emerged: ``little Higgs theories'' which realize the Higgs as a
pseudo-Goldstone boson in a low energy effective theory with a cutoff
parametrically above the weak scale of order $\Lambda \sim 10$ TeV.
In this paper and its companion we have presented two minimal models
of this sort. In the ``minimal moose'' model described here, two Higgs
doublets, a complex weak triplet and a complex scalar are the only
new degrees of freedom below the TeV scale.  At the TeV scale a set of
$SU(3)$ gauge bosons, an additional Higgs doublet, weak triplet and
singlet and colored fermion appear. These same-statistics partners of
the \SM\ fields are responsible for eliminating the 1-loop quadratic
sensitivity of the little Higgs masses to the UV physics at scale
$\Lambda$.

Our purpose in this paper and its companion has been to construct the
smallest examples of the little Higgs phenomenon, both for the sake of
economy as well as to illustrate the physics as clearly as possible.
There is still much left to explore, and many further issues to be
addressed by generalizations of these ideas. For example, perhaps the
most compelling argument for low-energy supersymmetry is the
spectacular prediction of the weak mixing angle with gauge coupling
unification not far from the Planck scale.  Recently it has been
realized that the weak mixing angle can also be correctly predicted by
electroweak unification into an $SU(3)$ symmetry at the $\sim 10$ TeV
scale \cite{Dimopoulos:2002mv}. 
Little Higgs models offer the possibility to implement this
mechanism in a natural way, and explicit
models of this type are currently being constructed.  

It is also of
obvious interest to construct UV completions of these models at scales
above $\Lambda$.  The non-linear sigma model can be completed into a
linear sigma model.  This leads to a new ``hierarchy problem'' at the
scale $10$ TeV, which can then be solved by supersymmetry broken at
$\sim 100$ TeV, alleviating nearly all the conventional constraints on
low-energy supersymmetry.  More daringly, the linear sigma model
fields themselves may arise as little Higgses in a larger theory,
perhaps extrapolating to extremely high energies in a ``cascade'' of
little Higgs models. 

We can also imagine UV completions where the link fields emerge from
fermion condensates in a strongly interacting gauge theory with a
strong scale $\sim \Lambda \sim 10$ TeV.  Since we have seen that all
the interactions required to produce the Higgs potential can be
triggered from the same couplings generating the top quark mass, the
main (and familiar) challenge is the implementation of the ETC-like
interactions needed to generate the Yukawa coupling operators.
However, the usual fatal flaws of strong dynamics at the TeV scale,
such as large corrections to precision electroweak observables and
too-light pseudo-Goldstone bosons, are eliminated in this framework,
since the scale of strong dynamics would be well above the TeV scale.
As we have seen the physics of electroweak symmetry breaking is still
perturbative with a light Higgs, and so precision electroweak
corrections are under control, and the lightest pseudo-Goldstone
bosons are the little Higgses themselves.  Furthermore, the difficulty
of generating flavor without excessively large FCNCs is also greatly
ameliorated \cite{Chivukula:2002ww}, and is a more tractable
model-building task.

The models presented here are new, fully realistic theories of TeV
physics with natural electroweak symmetry breaking. As such the
detailed phenomenology, constraints from precision low-energy
measurements, and the implications for present and future colliders
demand further exploration.

\acknowledgments
We would like to thank Sekhar Chivukula, Ken Lane and Martin Schmaltz for 
stimulating conversations.  
The work of A. Nelson and E. Katz was partially supported by the DOE under
contract DE-FGO3-96-ER40956.  N.A-H., T. Gregoire and J. Wacker are supported
in part by the Department of Energy under Contracts DE-AC03-76SF00098 and the
National Science Foundation under grant PHY-95-14797. A.G.C. is supported in
part by the Department of Energy under grant number DE-FG02-91ER-40676.
N.A.-H. is also supported by the Alfred P. Sloan foundation, and the David
and Lucille Packard Foundation. T. Gregoire is also supported by an NSERC 
fellowship.


\begin{thebibliography}{50}

\bibitem{Arkani-Hamed:2001nc}
N.~Arkani-Hamed, A.~G. Cohen, and H.~Georgi, {\it Electroweak symmetry breaking
  from dimensional deconstruction},  {\em Phys. Lett.} {\bf B513} (2001)
  232--240,
  [\href{http://xxx.lanl.gov/abs/http://arXiv.org/abs/hep-ph/0105239}{{\tt
  http://arXiv.org/abs/hep-ph/0105239}}].

\bibitem{Arkani-Hamed:2002pa}
N.~Arkani-Hamed, A.~G. Cohen, T.~Gregoire, and J.~G. Wacker, {\it Phenomenology
  of electroweak symmetry breaking from theory space},
  \href{http://xxx.lanl.gov/abs/http://arXiv.org/abs/hep-ph/0202089}{{\tt
  http://arXiv.org/abs/hep-ph/0202089}}.

\bibitem{Arkani-Hamed:2001ca}
N.~Arkani-Hamed, A.~G. Cohen, and H.~Georgi, {\it (de)constructing dimensions},
   {\em Phys. Rev. Lett.} {\bf 86} (2001) 4757--4761,
  [\href{http://xxx.lanl.gov/abs/http://arXiv.org/abs/hep-th/0104005}{{\tt
  http://arXiv.org/abs/hep-th/0104005}}].

\bibitem{Hill:2000mu}
C.~T. Hill, S.~Pokorski, and J.~Wang, {\it Gauge invariant effective lagrangian
  for kaluza-klein modes},  {\em Phys. Rev.} {\bf D64} (2001) 105005,
  [\href{http://xxx.lanl.gov/abs/http://arXiv.org/abs/hep-th/0104035}{{\tt
  http://arXiv.org/abs/hep-th/0104035}}].

\bibitem{Arkani-Hamed:2001ed}
N.~Arkani-Hamed, A.~G. Cohen, and H.~Georgi, {\it Twisted supersymmetry and the
  topology of theory space},
  \href{http://xxx.lanl.gov/abs/http://arXiv.org/abs/hep-th/0109082}{{\tt
  http://arXiv.org/abs/hep-th/0109082}}.

\bibitem{Georgi:1986hf}
H.~Georgi, {\it A tool kit for builders of composite models},  {\em Nucl.
  Phys.} {\bf B266} (1986) 274.

\bibitem{Douglas:1996sw}
M.~R. Douglas and G.~Moore, {\it D-branes, quivers, and ale instantons},
  \href{http://xxx.lanl.gov/abs/hep-th/9603167}{{\tt hep-th/9603167}}.

\bibitem{Gregoire:2002a}
T.~Gregoire and J.~Wacker, {\it Mooses, topology and higgs},
  \href{http://xxx.lanl.gov/abs/http://arXiv.org/abs/hep-ph/0206023}{{\tt
  http://arXiv.org/abs/hep-ph/0206023}}

\bibitem{Arkani-Hamed:2002a}
N.~Arkani-Hamed, A.~G. Cohen, E.~Katz, and A.~E. Nelson, {\it The littlest
  higgs},
  \href{http://xxx.lanl.gov/abs/http://arXiv.org/abs/hep-ph/0206???}{{\tt
  http://arXiv.org/abs/hep-ph/0206021}}

\bibitem{Lane:2002pe}
K.~Lane, {\it A case study in dimensional deconstruction},
  \href{http://xxx.lanl.gov/abs/http://arXiv.org/abs/hep-ph/0202093}{{\tt
  http://arXiv.org/abs/hep-ph/0202093}}.

\bibitem{Froggatt:1979nt}
C.~D. Froggatt and H.~B. Nielsen, {\it Hierarchy of quark masses, cabibbo
  angles and cp violation},  {\em Nucl. Phys.} {\bf B147} (1979) 277.



\bibitem{Dobrescu:1998nm}
B.~A. Dobrescu and C.~T. Hill, {\it Electroweak symmetry breaking via top
  condensation seesaw},  {\em Phys. Rev. Lett.} {\bf 81} (1998) 2634--2637,
  [\href{http://xxx.lanl.gov/abs/http://arXiv.org/abs/hep-ph/9712319}{{\tt
  http://arXiv.org/abs/hep-ph/9712319}}].

\bibitem{Chivukula:1998wd}
R.~S. Chivukula, B.~A. Dobrescu, H.~Georgi, and C.~T. Hill, {\it Top quark
  seesaw theory of electroweak symmetry breaking},  {\em Phys. Rev.} {\bf D59}
  (1999) 075003,
  [\href{http://xxx.lanl.gov/abs/http://arXiv.org/abs/hep-ph/9809470}{{\tt
  http://arXiv.org/abs/hep-ph/9809470}}].

\bibitem{Manohar:1984md}
A.~Manohar and H.~Georgi, {\it Chiral quarks and the nonrelativistic quark
  model},  {\em Nucl. Phys.} {\bf B234} (1984) 189.

\bibitem{Cohen:1997rt}
A.~G. Cohen, D.~B. Kaplan, and A.~E. Nelson, {\it Counting 4pi's in strongly
  coupled supersymmetry},  {\em Phys. Lett.} {\bf B412} (1997) 301--308,
  [\href{http://xxx.lanl.gov/abs/http://arXiv.org/abs/hep-ph/9706275}{{\tt
  http://arXiv.org/abs/hep-ph/9706275}}].

\bibitem{Luty:1998fk}
M.~A. Luty, {\it Naive dimensional analysis and supersymmetry},  {\em Phys.
  Rev.} {\bf D57} (1998) 1531--1538,
  [\href{http://xxx.lanl.gov/abs/http://arXiv.org/abs/hep-ph/9706235}{{\tt
  http://arXiv.org/abs/hep-ph/9706235}}].


\bibitem{Chivukula:2002ww}
R.~S. Chivukula, N.~Evans, and E.~H. Simmons, {\it Flavor physics and
  fine-tuning in theory space},
  \href{http://xxx.lanl.gov/abs/http://arXiv.org/abs/hep-ph/0204193}{{\tt
  http://arXiv.org/abs/hep-ph/0204193}}.

\bibitem{Katz:2002a}
E.~Katz and A.~E. Nelson, {\it in preparation}, .

\bibitem{Dimopoulos:2002mv}
S.~Dimopoulos and D.~E. Kaplan, {\it The weak mixing angle from an su(3)
  symmetry at a tev},
  \href{http://xxx.lanl.gov/abs/http://arXiv.org/abs/hep-ph/0201148}{{\tt
  http://arXiv.org/abs/hep-ph/0201148}}.

\end{thebibliography}
\providecommand{\href}[2]{#2}\begingroup\raggedright\endgroup

\end{document}